\documentclass[aip,graphicx,reprint]{revtex4-1}
\usepackage{graphicx}
\usepackage[caption=false]{subfig}
\usepackage{color}
\usepackage{bm}
\usepackage{multirow}
\usepackage{amsmath}
\draft 

\begin{document}

\def\gradB{ion $\boldsymbol{B}\times\nabla B$ }
\def\ExB{$\bf{E}\times\boldsymbol{B}$ }
\def\Lpar{$L_\textnormal{par}$ }
\def\c2plus{C$^{2+}$}
\def\m3{m$^{-3}$}

\title{Measuring the Electron Temperature and Identifying Plasma Detachment using Machine Learning and Spectroscopy}

\author{C.M. Samuell}
\email[]{samuell1@llnl.gov}

\affiliation{Lawrence Livermore National Laboratory, Livermore, California 94550, USA}

\author{A.G. Mclean}
\affiliation{Lawrence Livermore National Laboratory, Livermore, California 94550, USA}

\author{C.A. Johnson}
\affiliation{Auburn University, Auburn, Alabama 36849, USA}

\author{F. Glass}
\affiliation{General Atomics, San Diego, California 92186, USA}

\author{A.E. Jaervinen}
\affiliation{Lawrence Livermore National Laboratory, Livermore, California 94550, USA}

\date{\today}

\begin{abstract}
  
  A machine learning approach has been implemented to measure the electron temperature directly from the emission spectra of a tokamak plasma. This approach utilized a neural network (NN) trained on a dataset of 1865 time slices from operation of the DIII-D tokamak using extreme ultraviolet / vacuum ultraviolet (EUV/VUV) emission spectroscopy matched with high-accuracy divertor Thomson scattering measurements of the electron temperature, $T_e$. This NN is shown to be particularly good at predicting $T_e$ at low temperatures ($T_e < 10$ eV) where the NN demonstrated a mean average error of less than 1 eV. Trained to detect plasma detachment in the tokamak divertor, a NN classifier was able to correctly identify detached states ($T_e<5$ eV) with a 99\% accuracy (F$_1$ score of 0.96) at an acquisition rate $10\times$ faster than the Thomson scattering measurement. The performance of the model is understood by examining a set of 4800 theoretical spectra generated using collisional radiative modeling that was also used to predict the performance of a low-cost spectrometer viewing nitrogen emission in the visible wavelengths. These results provide a proof-of-principle that low-cost spectrometers leveraged with machine learning can be used both to boost the performance of more expensive diagnostics on fusion devices, and be used independently as a fast and accurate $T_e$ measurement and detachment classifier. \\

\end{abstract}

\maketitle

\section{Introduction}
\label{sec:introduction}

Using spectroscopy to measure plasma parameters including electron temperature, $T_e$, and density, $n_e$, has been a significant area of research in many areas of plasma physics for decades \cite{fantz2006, zhu2010}. The light emitted by excited ions in particular, has been studied extensively as their brightness is a strong function of $T_e$ and $n_e$. Their emission can be measured passively, non-invasively, and distanced from the plasma source, making it an attractive technique for fusion plasmas. For measurement of $T_e$, the most common approach is finding ratios of several of these emission lines, removeing the density dependence and thereby isolating the temperature dependence. These line ratios are generally discovered and profiled using detailed theoretical models, such as collisional radiative models, that use atomic physics data to predict line brightness for a given ion species. In practice, suitable line combinations are rare, and those that are discovered can ultimately be difficult to observe experimentally or are contaminated by other lines. In some instances, the addition of impurity gases to the plasma is required, further complicating the experimental setup, and risking perturbation, although it does provide the advantage of measurement localization \cite{schmitz2008, barbui2019, griener2017}. On the {DIII-D} tokamak and other carbon-walled machines, there has been significant effort to make use of carbon line ratios for $T_e$ measurement \cite{isler1997, vlad2016}. The measured temperature from line-ratio techniques is often reported as an `effective' temperature, denoting that the measurement is weighted by the location of the emitting species along the viewchord line of sight rather than the temperature at a particular location. For example, a measurement solely using C$^{2+}$ lines is unlikely to be accurate at low or high temperatures where C$^{2+}$ density is low, will be weighted towards the region of highest C$^{2+}$ emission brightness, and the measurement location will move as the region of bright C$^{2+}$ emission moves with changing temperature spatial profiles. Conversely, Thomson scattering, a laser-based diagnostic used commonly for temperature measurement in the core and pedestal regions of many tokamaks, provides very accurate measurement of the temperature although it is a complex and often-costly system requiring high-powered lasers and access to multiple vessel ports \cite{carlstrom1992}. Thomson scattering is a point-measurement and as a result does not suffer the same interpretation problems that can come from the need to identify the emission location as is the case in spectroscopic measurement of $T_e$. Supervised machine learning methods offer a way that the accuracy of Thomson scattering can be leveraged to improve spectroscopic measurements of $T_e$ by providing a set of `correct' reference labels for spectrometer data. Thus the speed and ease of spectroscopy is combined with the accuracy of the Thomson scattering, providing a spectroscopic temperature measurement that does not rely solely on theoretical underpinnings.\\

Machine learning has found increasing use in a wide range of applications in the physical sciences \cite{mehta2019, spears2018} due to its ability to leverage large datasets and to capture the complex nonlinear and interdependent relationships common in physics. For fusion plasmas, it has recently been employed for real-time prediction of disruption events \cite{rea2019}, enhancing the tomographic inversions of a bolometer diagnostic \cite{matos2017}, and to provide speed improvements to computationally intensive theory-based models \cite{meneghini2017}. This work applies recent developments in machine learning techniques to the long-studied problem of measuring $T_e$ using spectroscopy. Utilizing supervised learning, particularly neural networks, to measure $T_e$ has been achieved in several other studies. For example, soft-x-ray arrays combined with Thomson scattering have been used to predict core $T_e$ profiles on the NSTX spherical tokamak\cite{clayton2013} and measure $T_e$ using Thomson scattering directly \cite{lee2016} on the KSTAR tokamak. This is the first work to apply neural networks to measure $T_e$ in the divertor plasma, and to use a collisional radiative model to interpret the neural network performance.\\

The underlying dataset linking spectrometer and Thomson scattering measurements is discussed in Section \ref{sec:database} followed in Section \ref{sec:DNN} by the experimental demonstration of a neural network trained to measure $T_e$. The experimental results are examined and extrapolated to future diagnostics using a neural network trained on simulated data in Section \ref{sec:colradpy}. Finally, the limitations of the technique and directions for future work are outlined in Section \ref{sec:conclusion}. \\

\begin{figure*}[t]
    \centering
    \subfloat{\includegraphics[width=1\textwidth]{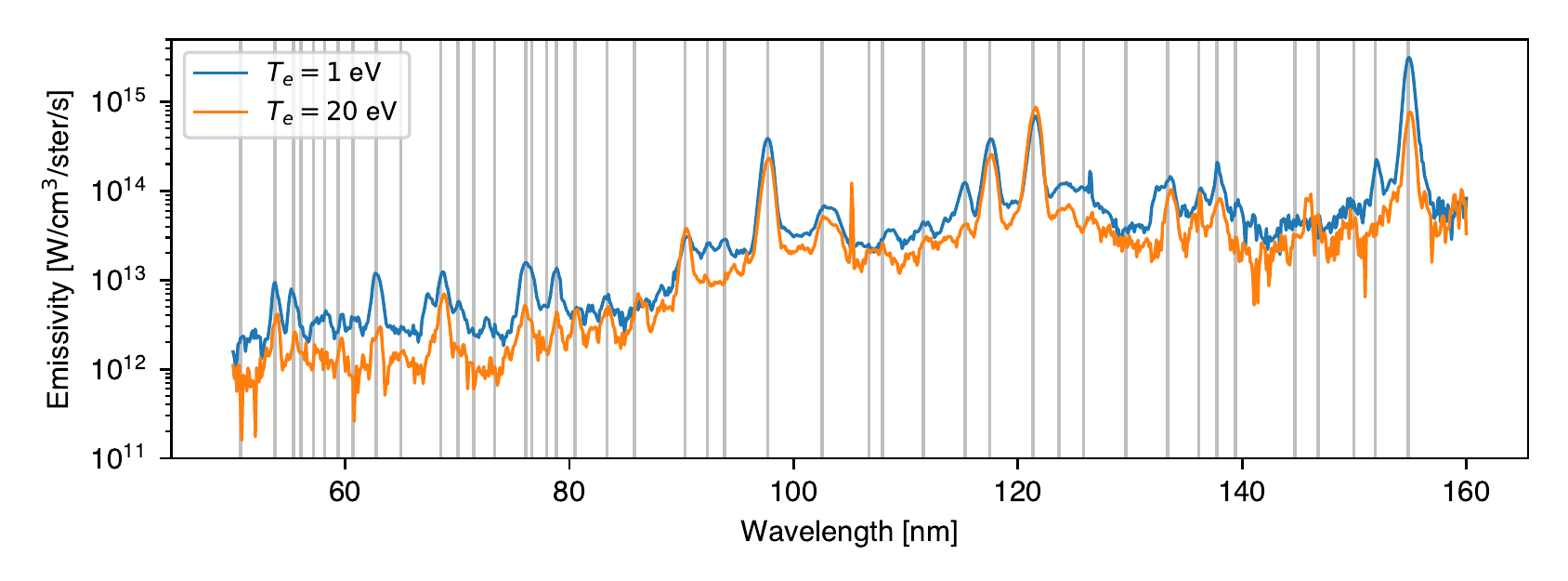}}
    \caption{\label{fig:spectra_example} Example DivSPRED EUV/VUV spectra at two differing DTS-measured electron temperatures with approximately the same electron density ($\sim 1 \times 10^{19}$ m$^{-3}$). Vertical gray lines indicate the location of 44 distinct emission lines identified in the experimental dataset. }
\end{figure*}

\section{The Experimental Dataset}
\label{sec:database}

To test machine learning (ML) models for spectroscopy, a dataset was established matching line-of-sight integrated spectra from a EUV/VUV spectrometer with localized measurements of the electron temperature made using the divertor Thomson scattering on DIII-D. Each dataset entry is a single time slice in a DIII-D discharge from the 2018 and 2019 run campaigns (DIII-D shots 173048 - 177059). The construction and cleaning of this dataset is the most important contributor to the quality of the subsequent machine learning models and their underlying assumptions, and so this process is described in detail below. \\

\begin{figure}[t]
    \centering
    \subfloat{\includegraphics[width=1\columnwidth]{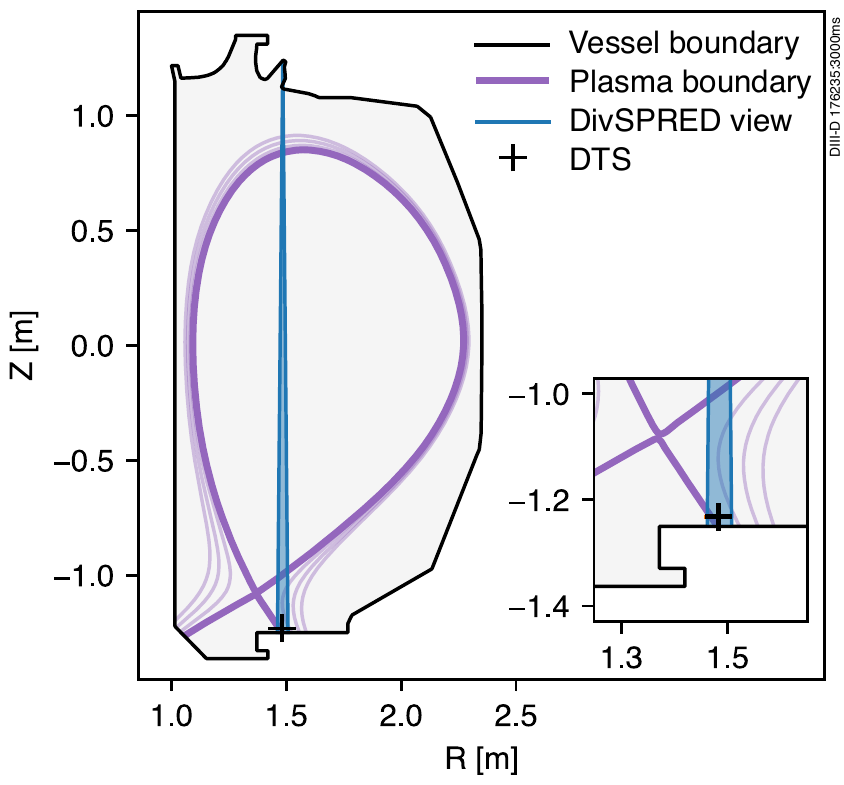}}
    \caption{\label{fig:efit} Cross-section of the DIII-D vessel (black) and plasma as defined by the last closed flux surface and magnetic field lines (purple). The measurement location for the single DTS channel used for this study ($T_e$, black $+$) is contained within the DivSPRED sprectrometer field of view (blue). Note that the majority of light observed by the spectrometer comes from the divertor area below the X point at $Z\sim -1.1$ m. Inset shows a zoomed view. }
\end{figure}

The plasma emission spectra were sourced from the absolutely-calibrated DivSPRED EUV/VUV spectrometer \cite{mclean2019tech, mclean2017iaea}. This spectrometer covers wavelengths from around 50 - 160 nm with a full-width, half-maximum instrument width of 1 nm. Reference spectra at two different electron temperatures are shown in Figure \ref{fig:spectra_example}. The spectrometer is vacuum-coupled to the DIII-D vessel via an upper vertical port, viewing through the `core' plasma and into the divertor. The majority of emitted light originates in the divertor, a result of the low electron temperatures there ($<100$ eV) relative to the core plasma (100 - $\sim$8000 eV). This viewing geometry is shown in Figure \ref{fig:efit} for a typical plasma shape. While DivSPRED can operate at acquisition rates above 1kHz, for the data presented here, it was nominally operated at a rate of 545 Hz with each time slice integrated over an acquisition period of 1.8 ms. \\

Electron temperature measurements were made with the Divertor Thompson Scattering (DTS) system \cite{carlstrom1997}. While the full DTS system makes several measurements at varying heights in the divertor, the presented results focus on a single measurement location 2.9 cm above the divertor plate as shown in Figure \ref{fig:efit}. This DTS measurement samples a roughly cylindrical volume about 12 mm tall and 5 mm wide at a rate of 50 Hz and integration time of about 15 ns.\\

Fundamentally, the spectrometer and DTS make very different kinds of measurement. Where DTS is a very localized measurement in time and space, the spectrometer integrates over a time period five orders of magnitude longer and collects light from a larger volume; two parameters over which the plasma conditions may vary considerably. To reduce the impact of the spatial difference, the dataset is limited to time slices where the magnetic field strike point is close to, or radially inboard of, the spectrometer line of sight and a DTS channel close to the target is chosen.  Additionally, the difference in acquisition time between the two diagnostics makes DTS sensitive to fast fluctuations that are not accessible to the spectrometer. A single data point or a small-sample dataset may therefore better reflect a sampling of transient events and local turbulence rather than steady-state conditions. This highlights the need for a large dataset of measurements and the underlying assumption that the divertor turbulence will result in statistical fluctuations around the mean value sampled by the slower spectrometer diagnostic.  \\

Several data processing utilities were combined to generate the dataset. The initial shot survey was performed using SQL queries of the D3DRDB relational database of discharge overview data \cite{schissel2000}. Subsequently, the TokSearch Apache Spark utility was used to collect time-series data and align time bases of the spectra with DTS measurements and the metadata used for shot/time down-selection \cite{sammuli2018}. Candidate time slices were down-selected aggressively according to the following set of requirements:
\begin{itemize}
\item Discharge shape must be a lower single null (LSN) configuration with the strike point at a major radius between 1.47 - 1.50 m. This location places the strike point near the spectrometer and DTS measurement locations.
\item The plasma shot type must be from the regular scientific campaign; testing and development shots are excluded.
\item Spectrometer pixels must be well lit and DTS temperatures must be non-null.
\item Times before 2000 ms and above 5000 ms are excluded to reduce the possibility of off-normal events (e.g. end-of-shot plasma current rampdown) and the fast-changing development portion of discharges (e.g., ramping input power at the beginning of a discharge).
\item Discharges with measurable quantities of nitrogen or neon emissions were omitted with impurity puffing preferentially used in detached time slices, providing a proxy for low-temperatures for the ML model. The model could learn this proxy which would not be generalizable to new data that may have the presence of impurities at any temperature. This was quantified using the emission lines detected by the core SPRED spectrometer \cite{fonck1982}.
\item Shots with fewer than 5 time slices that satisfy the above criteria are rejected to reduce the likelihood of outliers.
\end{itemize}

\begin{figure*}[t]
    \centering
    \subfloat{\includegraphics[width=1\textwidth]{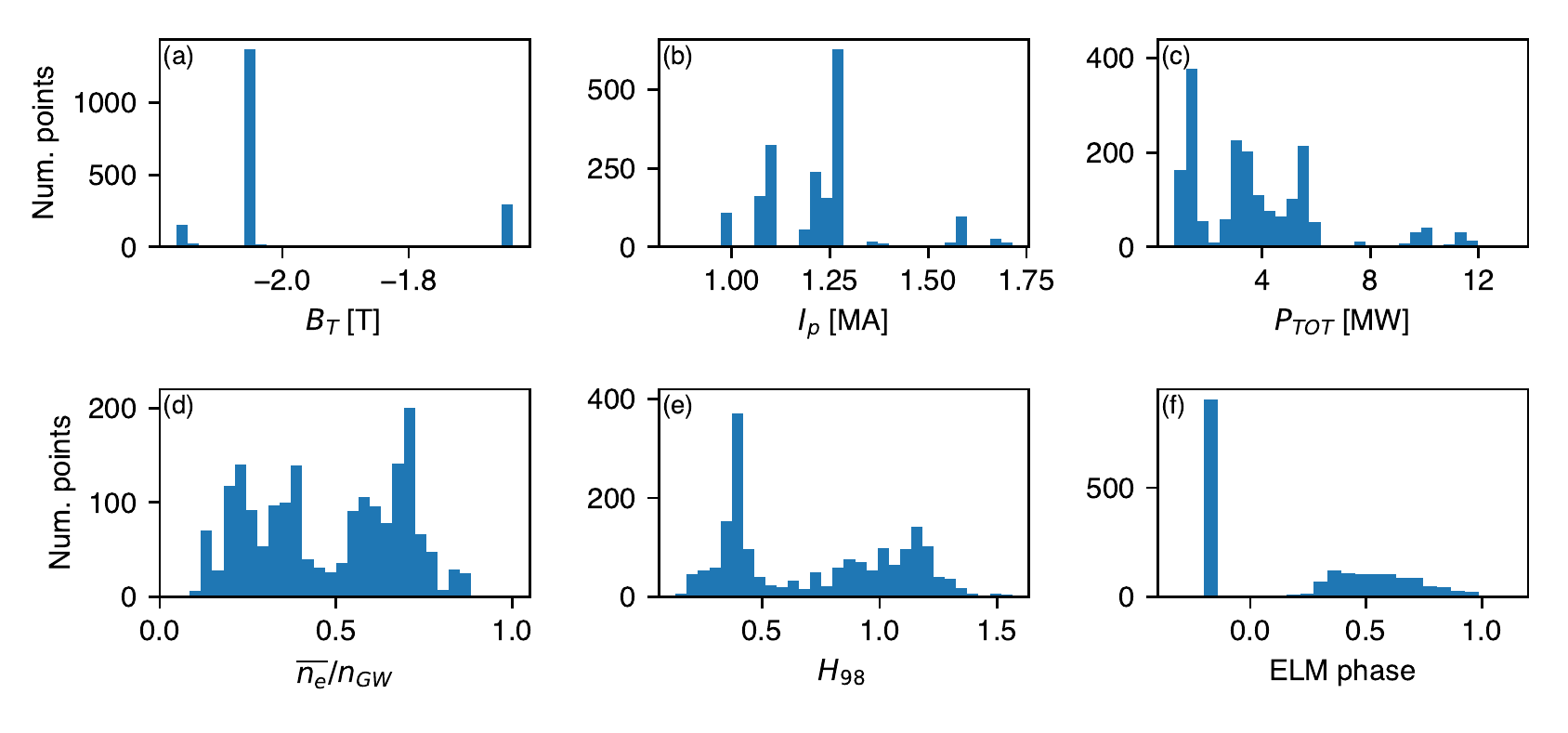}}
    \caption{\label{fig:database_metadata} The distribution of discharge parameters among time slices included in the database including (a) the toroidal field strength, (b) the plasma current, (c) the total input power to the discharge, (d) a normalized line-averaged density, (e) the normalized plasma confinement, and (f) the ELM phase where a value of -0.2 indicates L-mode (no ELMs).}
\end{figure*}

Additionally, for shots that had at least some portion of H-mode confinement, the presence of fast-transient ELM events had to be considered. In these shots, the ELMs were identified and time-slices were categorized as either inter- or intra-ELM using ELM processing workflows inside the OMFIT framework \cite{meneghini2015}. The ELM phase is defined as 0 at the end of an ELM rising to 1 at the beginning of the following ELM event. For this dataset, time slices were rejected if the ELM phase was below 0.3 or above 0.95. Unlike DTS which could be treated as an instantaneous measurement, the spectrometer acquires data over an integration time on the same timescale as the ELM-events. For this case, and for other finite acquisition-length diagnostics (typically those diagnostics using cameras), the ELM-phase had to be between 0.3 and 0.95 for the duration of the camera integration time. This filtering allows the dataset to include both L-mode and H-mode discharges. \\

It should be noted that the methodology for constructing the database outlined here is inherently biased against some DIII-D plasma operating regimes. For example, grassy-ELM regimes \cite{ozeki1990}, which have high-frequency ELM events will have few quiescent periods long enough for an unaffected spectrometer acquisition to be found. Some detachment scenarios where there has been significant pedestal degradation or those operating very close to the H-L mode back-transition can also display high-frequency pedestal and divertor oscillations. This affects the applicability of the as-built model to these regimes but is an effect that could be ameliorated with future DIII-D experiments generating data in these regimes expressly to provide a training dataset for ML applications. \\

To give a high-level overview of the database's scope, the distribution of time slices across a range of plasma parameters are shown in Figure \ref{fig:database_metadata}. The toroidal field, $B_T$, shows only negative values corresponding to the `forward' toroidal field direction on DIII-D (\gradB into the active divertor for lower-single null plasmas) and only at a selection of nominal values. The dataset would be more complete if both directions could be included as the action of \ExB drifts (which change direction when the toroidal field changes direction) can substantially alter the distribution of electron temperatures and densities in the divertor \cite{rognlien1999b,rognlien2017}. However, amongst the candidate time slices in the `reversed' field direction, $\sim$ 99 \% had $T_e > 10$ eV resulting in a very biased subset of the dataset. It is temperatures less than 10 eV that are most important for the detached conditions impacted strongly by this \ExB effect. To avoid introducing this bias, reversed toroidal field time slices were omitted reducing the dataset size and introducing a model assumption that \gradB drift is directed into the active divertor. This is a consequence of the particular experiments run during the selected run periods rather than an underlying feature of DIII-D plasmas. \\

A range of plasma currents, $I_p$, and total inputted powers $P_{TOT}$, are represented with conditions that are peaked around $I_p = 1.25$ MA and $P_{TOT} < 6$ MW. $P_{TOT}$ is defined as the input power to the plasma comprising neutral beam injection (NBI), electron cyclotron heating (ECH), and an ohmic contribution. The majority of the shots in this database are dominantly NBI heated. The Greenwald fraction ($n_{GW} \equiv \frac{I_p}{\pi a^2}$ where $a$ is the minor radius \cite{greenwald2002}) is shown to give an indication of the range of line-averaged densities in a form that can be conveniently compared to other tokamaks. The database contains Greenwald fractions between $0.08$ and $1.04$ corresponding to line-averaged electron densities of $\overline n_e$ between $9.6\times 10^{18}$ and  $1.6\times 10^{20}$ m$^{-3}$. The distribution peaks occur around $n_e / n_{GW} = 0.2$ and $0.7$ corresponding to common operating regimes. A measure of plasma confinement is given by the unitless normalized confinement factor, $H_{98}$, as referenced to the ITER physics basis scaling IPB98(y,2) \cite{iter1999} for which  $H_{98} \gtrsim 1$ is typical of H-mode confinement and  $H_{98} \ll 1$ is typical of L-mode confinement. In the dataset, about 30\% of time slices have $H_{98} > 1$. Finally, Figure \ref{fig:database_metadata} (f) shows the ELM phase where an ELM phase of -0.2 corresponds to a low-confinement (L-mode) discharge without ELMs. \\

\begin{figure}[t]
    \centering
    \subfloat{\includegraphics[width=0.9\columnwidth]{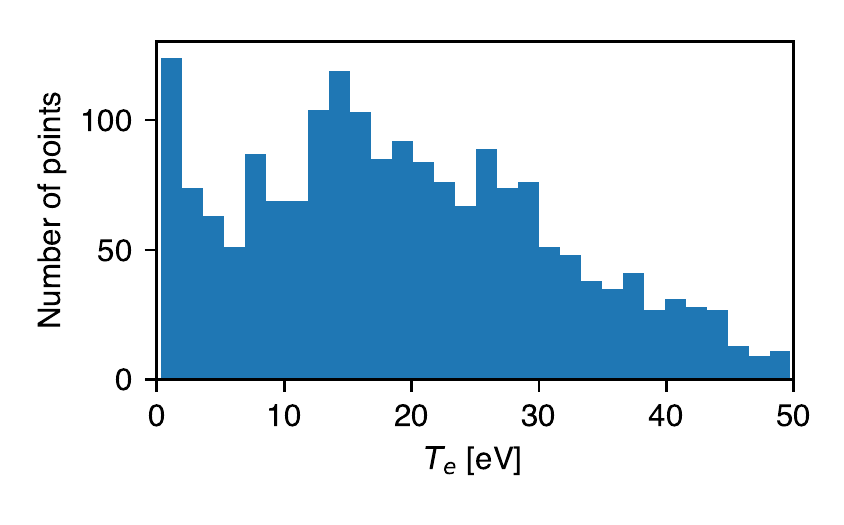}}
    \caption{\label{fig:Te}Distribution of electron temperatures in full dataset displaying frequency peaks in both the attached and detached plasma regimes.}
\end{figure}

The resulting database contains 1865 time slices from 98 discharges. The distribution of $T_e$ values in the dataset are shown in Figure \ref{fig:Te}. Temperatures range from 0.3 eV to a cap of 50 eV chosen to reduce high-temperature outliers that tend to result from pedestal fluctuations and non-filtered ELMs. Two distinct peaks in conditions are present that could be generally described as `attached' (20 eV) and `detached' (1 eV) representing common areas of interest for experiments from the run period. While precise definitions of detachment are challenging, about 14\% of the database is detached using $T_e<5$ eV as a detachment threshold. \\

For training ML models, the database was shuffled and split into three subsets. 25\% of the data is a holdout set that is reserved for independent testing of the model performance. This data is never seen by the model during training so that it can be used to assess the performance of the model on new data. Of the remaining 75\% of the data, 25\% is taken for testing the model performance between training cycles with the remaining data used directly for model training. For this dataset, this equates to 1046 points used for training, 355 points used for assessing model training progress and comparing models, and a 467 point evaluation test set. \\

\section{A Neural Network Trained on Experimental Data}
\label{sec:DNN}

A feed-forward neural network (NN) is a supervised learning technique by which a series of nested non-linear functions are adjusted iteratively to fit a dataset \cite{spears2018}. For a regression task, such as predicting a continuous variable given some input, the NN provides a function that transforms the input data to an output quantity of interest (e.g. transforming emission spectra to an electron temperature). The NN has the ability to capture substantial non-linearity, particularly for so-called `deep learning' models with many layers and subsequently many free parameters. The high number of free parameters makes the NN able to support enormous complexity but also makes it difficult to train; large datasets are required to train these models and the training process itself is prone to numerical instability. Care must be taken that the training process results in a generalizable model rather than overfitting by which each detail of the training data is `memorized' by the model leaving it prone to large errors when faced with previously unseen data.\\

\begin{figure}[t]
    \centering
    \subfloat{\includegraphics[width=0.9\columnwidth]{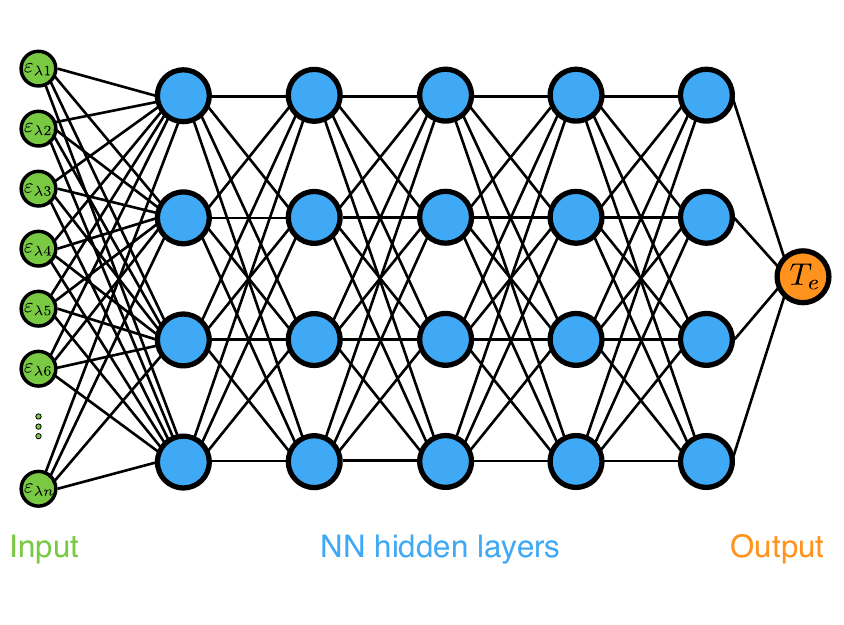}}
    \caption{\label{fig:nn_schematic} A neural network schematic for an example model five layers deep (horizontal axis) and four layers wide (vertical axis) that takes a single spectra as input and transforms it to a prediction on the electron temperature.}
\end{figure}

The schematic for a NN is shown in Figure \ref{fig:nn_schematic}. The input-layer receives the data and is followed by a number of layers of multiple neurons before reaching the output layer. The layers between the input and output are often referred to as `hidden' layers. This example NN is `fully-connected' such that each neuron is connected to each neuron in the previous layer and each neuron in the following layer. Following the description by Spears \emph{et al.} \cite{spears2018}, each individual neuron takes as inputs the output from each neuron in the previous layer and act upon those inputs with a non-linear activation function before passing along the new value to each neuron in the subsequent layer. The net effect is a nonlinear transformation at each layer with the final output being the result of the nested set of functions from each layer acting upon the results of the previous one. The relative weighting and bias of each neuron are the adjustable parameters that are modified during the model training process which optimizes by minimizing some specified loss function (e.g. the mean squared error). The training of neural networks is similar to traditional regression techniques whereby a line may be fitted to some data by varying a set of variables optimized by minimizing the least-squares difference between the line and the data. In practice, a dataset of matched inputs and outputs is supplied to the model in small batches, passing the data from the input layer through the hidden layers and to the output. The loss function is then calculated to quantify the distance between the model's predictions and the correct values. The error gradient is then propagated backward through the model to provide updated values to the model's weights and biases. The full training process involves moving through the training dataset multiple times to incrementally improve the model's performance with respect to the loss function. \\

\subsection{Prediction of the electron temperature}

A NN was trained to predict the electron temperature at a single point using the database of EUV/VUV spectrum labeled with Thomson scattering-measured $T_e$ as described in Section \ref{sec:database}. The precise architecture for the model (number of hidden layers, number of neurons per layer, activation function, etc.) was chosen by testing many architectures and comparing their performance (i.e. hyperparameter tuning). The best performing model was a NN with 12 hidden layers of 12 neurons that were sandwiched between a 1000 element input vector (the spectra) and the single element output vector ($T_e)$. The resulting model contained 13,741 trainable parameters. The input vector is the absolutely calibrated intensity of light at a single slice intepolated onto a 1000 pixel array between 50-160 nm to provide a standard wavelength for each pixel irrespective of the precise grating alignment which can change between spectrometer calibrations. The output is the electron temperature in eV. The dataset of spectra is scaled by removing the mean and scaling to a variance of 1 for each wavelength. Each neuron in the model used an exponential linear unit (ELU) activation function with the exception of the final output neuron which does not have an activation function so that it can take on any value. Activation function choice is important for numeric stability during the training process and convergence speed. Model weights were initialized using He initialization \cite{he2015} and updated at a learning rate that decreased exponentially every 30 epochs (i.e. every 30 passes through the training set). A Nadam optimizer was used to calculate the changes to the model weights \cite{dozat2016}. Among other improvements on the classic stochastic gradient descent approach to deriving these weights, Nadam includes a momentum term to account for the speed at which the loss function gradient had been changing in recent updates that were administered in previous training epochs as calculated at a point in the direction that the momentum term was pointing \cite{geron2019}. The training process was deemed completed after the model performance hadn't improved in 50 training epochs (up to a maximum of 1000 epochs) resulting in an 89 epoch training cycle for this model. The mean absolute error was used as the loss function which offered more stability during the training process compared to the mean squared error. To avoid over-fitting, the model was penalized by adding $L_2$ regularization that resulted in poorer training-set performance but convergence to an improved performance on the evaluation dataset. The model construction and training were handled with the Tensorflow Python package \cite{tensorflow} trained either on a standard laptop CPU or using a NVIDIA Tesla V100 GPU for larger models and for the computation-intensive hyperparameter tuning. \\

\begin{figure}[t]
    \centering
    \subfloat{\includegraphics[width=1\columnwidth]{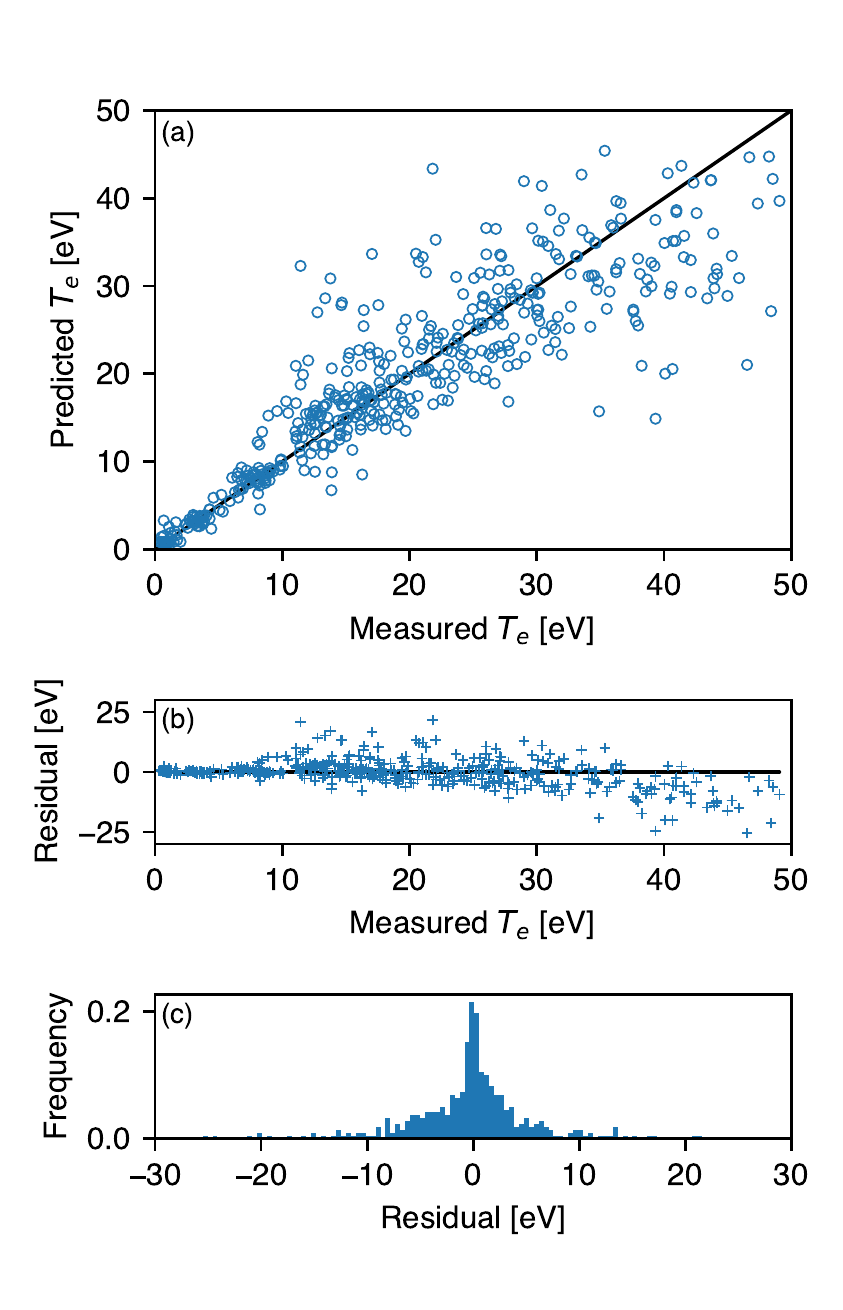}}
    \caption{\label{fig:nn_performance} Performance of the neural network model on the evaluation test set. The predicted temperatures as a function of the actual temperatures are shown in (a) alongside the residuals (predicted - measured) as a function of the measured temperature (b), and the distribution of those residuals (c).}
\end{figure}

The model's performance is evaluated by examining the 467 time slices that were withheld from the model training. This performance is shown in Figure \ref{fig:nn_performance} where the model's predicted $T_e$ is compared to the real (measured) $T_e$. For $T_e$ under 10 eV, the mean absolute residual was 0.8 eV with a standard deviation, $\sigma$, of 1.1 eV. For $T_e$ above 10 eV, the spread in predicted values grows with increasing temperature and the residuals depart from a mean closer to zero for $T_e \gtrsim 35$ eV indicating a systematic error at the highest temperatures. For the entire test dataset, the mean absolute residual was 3.6 eV ($\sigma = 4.3$ eV). This indicates that the region of highest performance for is the temperature range of most interest for detachment studies ($<10$ eV) where the accuracy is sufficient for a broad range of divertor physics studies. \\

\begin{figure}[t]
    \centering
    \subfloat{\includegraphics[width=1\columnwidth]{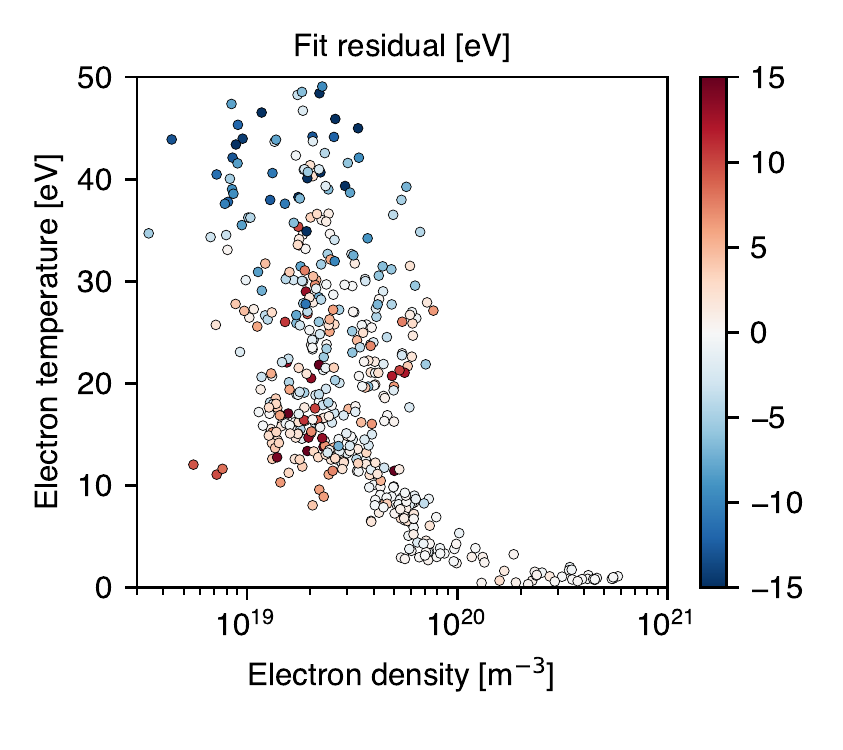}}
    \caption{\label{fig:nete_residuals}The difference between real and predicted values as a function of experimentally measured density and temperature showing how model performance changes as a function of plasma conditions. Note that the color map saturates at $\pm 15$ eV; the full residual range can be seen in figure \ref{fig:nn_performance}b.}
\end{figure}

The dependence of model performance on the plasma's density and temperature is shown in Figure \ref{fig:nete_residuals} with $n_e$ measured using DTS at the same location as the $T_e$ measurement. The strong link between low temperatures and high densities results from the absence of impurity-assisted detachment shots in the dataset. In these discharges, detachment is often achieved by increasing the fueling, most of which becomes recycling flux in the divertor; as the density rises, the temperature tends to drop and the plasma detaches. This parameter space could be expanded in the future with designated experiments examining detachment at higher power, increasing impurity concentrations, and at other plasma currents. The residual tends to be larger at higher temperatures (and low density) with larger residuals tending to be negative at high temperature and positive at moderate temperature. This trend can also be observed in Figure \ref{fig:nn_performance}(b) and is further discussed in Section \ref{sec:colradpy} where it is compared to the results for simulated spectra. \\

\begin{figure}[t]
    \centering
    \subfloat{\includegraphics[width=1\columnwidth]{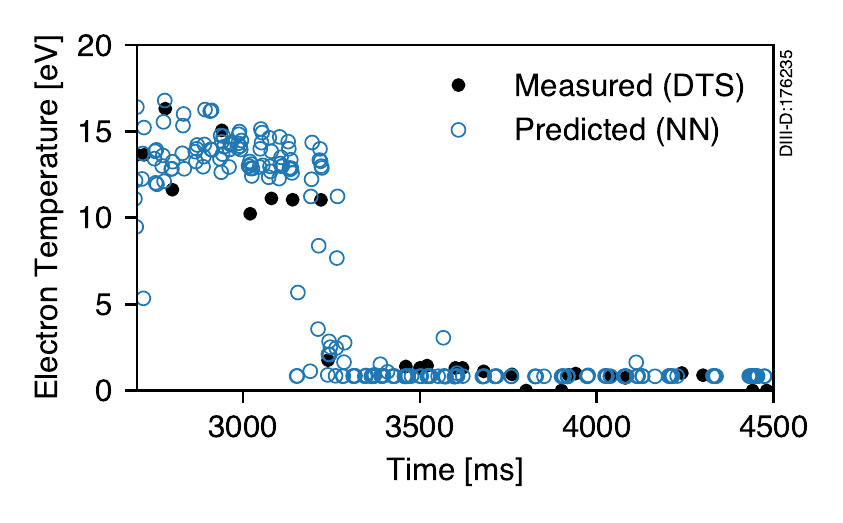}}
    \caption{\label{fig:single_shot} Electron temperature predicted with the NN model compared to the experimentally measured DTS value on a plasma discharge that the model had not seen during training. The rapid drop in $T_e$ occurs due to the onset of detachment. }
\end{figure}

An important separate validation exercise is to test the model against a plasma shot that was held-out in its entirety. The test, validation, and training datasets are sourced from a single randomized pool of data points such that time slices from a single shot may appear in each subset although no two time slices are identical and so an entirely separate shot provides a unique test. The shot chosen was a well-studied H-mode discharge with a density ramp that causes a detachment transition during the shot and subsequent transition from high $T_e$ to low $T_e$. This particular plasma was nearing the transition between a high- and low- confinement state (H-mode to L-mode) and so as the density rises, the ELM frequency increases substantially meaning that many data-points that were not between an ELM phase of 0.3 and 0.9 were removed due to ELM contamination. The comparison between measured and NN-predicted $T_e$ values is shown in Figure \ref{fig:single_shot}. Uncertainty on DTS $T_e$ measurements can be very small ($<0.3$ eV) however statistical spread due to the sampling of underlying turbulence not seen by the spectrometer can be several eV. This shot highlights the substantial improvement in the temporal performance of the spectroscopy+NN system compared to DTS. The acquisition speed of the spectrometer is $>10\times$ that of DTS in this discharge resulting in significantly more valid data points after ELM-filtering. This temporal resolution is particularly useful during the fast transition to detachment around 3230 ms. Whereas DTS indicates a stepwise transition between 11 eV and 2 eV, the NN approach indicates that there may be intermediate-valued temperatures between those two DTS points and dithering between attached and detached states as early as 3150 ms. This suggests that the transition from attached to detached may be more continuous than DTS is able to measure, a feature that would be important for our understanding of \ExB driven detachment \cite{jaervinen2018}. \\

\subsection{Detachment Classifier}

Instead of using the NN to predict $T_e$, a NN can instead be used to identify detachment conditions directly leveraging the high accuracy of the model observed for low temperatures. An accurate detachment indicator is useful for identifying the fueling or impurity level that induced detachment for comparisons between divertor types. Historically, the roll-over in ion saturation current measured by Langmuir probes, or the $T_e$ measured by Thomson scattering is used, however, both techniques have their disadvantages. Here we adopt $T_e < 5$ eV as a definition of detachment noting that this is the temperature below which ion-neutral elastic collisions and charge-exchange dominate, and electron pressure drops in the divertor due to local dissipation \cite{fenstermacher1997, loarte1998, leonard2018, boedo2018}.\\
 
The NN configuration is the same as was used for regression with the exception that the output neuron is replaced with a neuron with a sigmoid activation and the model retrained as a classifier. In this case, the NN is learning to predict a binary state (attached or detached) and outputting the probability that the time slice of interest is detached. The model is a binary classifier trained using a log-loss function (i.e the binary cross entropy) as the loss value to be minimized during model training. While the overall architecture of the model is almost identical to the one used for $T_e$ regression, the retraining allows the classifier to optimize to its specific task. The NN trained in this way achieved an accuracy of 99\% at classifying time slices as attached or detached in the evaluation dataset. \\

About 14\% of the dataset represents detached cases implying a skewing towards attached cases that makes accuracy a somewhat misleading indicator of model quality. For example, a model that predicted attached states 100\% of the time would still be 86\% accurate on this dataset. The $F_1$ score is a better measure of the classification performance which takes into account the false positives ($FP$) and false negatives ($FN$) alongside the true positives ($TP$). The $F_1$ score is the harmonic mean of the model's precision ($\frac{TP}{TP + FP}$) and recall ($\frac{TP}{TP + FN}$) such that:
\begin{equation}
  F_1 = 2\left( \frac{precision \times recall}{precision + recall}\right) = \frac{TP}{TP + \frac{1}{2}\left(FP + FN\right)}.
\end{equation}
A $F_1$ score of 1 would indicate perfect performance. The detachment classifier described here displayed a $F_1$ score of 0.96. The confusion matrix that breaks down the model's performance into constituent true and falsely predicted values are shown in Table \ref{tab:confusion}. Of the 467 cases in the evaluation dataset, 462 values cases were correctly labeled and 5 cases were incorrectly labeled. \\

Since the direct output of the NN can be taken as the probability that a case is attached, the degree to which the NN is certain of its predictions can be evaluated. An assessment of the model's certainty about a particular prediction gives a mechanism by which the time slices for which the model doubts its predictions can be discarded or flagged for manual review to further increase the diagnostic's performance. The vast majority of the model predictions are very confident; 95.5\% of the time slices in the test set were predicted with classification probabilities above 99.5\%. By removing the 4.5\% of that were less than 99.5\% certain (21 cases), the model predictions of attached and non-detached cases become 100\% correct (F$_1$ = 1). This removal of 4.5\% of the dataset to achieve a perfectly performing classifier is only a small reduction overall given the $10\times$ increase in sampling rate that the system achieves relative to DTS. \\

\begin{table}
  \centering
  \begin{tabular}{cc|cc}
    \multicolumn{1}{c}{} &\multicolumn{1}{c}{} &\multicolumn{2}{c}{\textbf{Predicted} \vspace{0.2cm}} \\ 
    \multicolumn{1}{c}{} & 
    \multicolumn{1}{c|}{} & 
    \multicolumn{1}{c}{Detached} & 
    \multicolumn{1}{c}{Attached} \\[1.5ex] \hline 
    \multirow[c]{2}{*}{\rotatebox[origin=tr]{90}{\textbf{Actual}}}
    & \hspace{0.2cm} Detached \hspace{0.2cm} & TP: 57 & FN: 2   \\[2.5ex]
    & \hspace{0.2cm}Attached \hspace{0.2cm} & FP: 3  & TN: 405 \\
  \end{tabular}
    \caption{\label{tab:confusion}A confusion matrix comparing the number of true positive ($TP$), false positive ($FP$), true negative ($TN$), and false negative ($FN$) cases detected by the detachment classifier on a previously unseen test dataset of 467 values.}
\end{table}

\section{Comparison with Simulated Data}
\label{sec:colradpy}

To further investigate NN performance for predicting the temperature from spectroscopy data, a database of simulated spectra was generated using collisional radiative (CR) physics modeling. The spectra and dataset were designed to emulate the characteristics of the DivSPRED spectrometer and experimental dataset so that the NN performance could be compared. The CR model balances the rate equations for populating and depopulating the ion's ground, metastable, and excited states taking into account \cite{johnson2019, wunderlich2009}:
\begin{itemize}
\item electron impact excitation and de-excitation,
\item electron impact ionization,
\item spontaneous emission,
\item recombination (radiative, dielectronic, and three-body), and
\item charge exchange.
\end{itemize}

For the simulations used here, the recently developed ColRadPy CR model was used \cite{johnson2019}. Simulations include the atomic processes that define the electronic states and energy transitions between states via electron impact, spontaneous emission, recombination, and charge exchange, and so capture the dominant physics mechanisms that define the atomic spectra. However, the simulated spectra will not \emph{exactly} reproduce experimental ones as transport effects, spin-state resolution ($J$ splitting), the presence of multiple species, line-broadening/splitting from ion temperature and magnetic field, and the contribution from molecular states are not considered.\\

\subsection{A Neural Network Trained on Simulated Carbon Spectra in the EUV/VUV Wavelengths}

The simulated spectra dataset consists of 4800 spectra with $1\times 10^{18} < n_e < 1\times10^{21}$ m$^{-3}$ and $0.5 < T_e < 50 $ eV. Density and temperature ranges were chosen to broadly cover the range of values expected experimentally on DIII-D. Individual density-temperature pairs were chosen with a uniformly random distribution in linear-space for $T_e$ and log-space for $n_e$ to account for the several orders of magnitude over which density meaningfully varies in the divertor plasma. In practice, not all combinations of density and temperature are routinely accessed on DIII-D; conservation of pressure dictates that high density is normally correlated with low temperature and vice-versa. Some combinations of density and temperature would represent extreme scenarios or inaccessible regions on DIII-D (high density and high temperature for example). \\
     
The DivSPRED emission spectra are vastly dictated by carbon emission as carbon is the dominant impurity on DIII-D owing to its graphite walls. A notable exception is emission from neutral deuterium emission, particularly Lyman-alpha. Deuterium is the fuel species, however, deuterium ions do not have a bound electron and consequently ionized deuterium does not produce line emission. Inclusion of both deuterium and carbon spectra in the simulated database would require accounting for a variable impurity concentration which would substantially complicate interpretation. Isolation of the carbon contribution allows us to focus on the role of atomic carbon emissivity for predicting $T_e$ noting that in an experimental setting there is other information that may be useful (continuum, molecules, etc.) \\

The simulated spectra are constructed by adding lines from C neutrals as well as C$^{+}$, C$^{2+}$, C$^{3+}$, C$^{4+}$, C$^{5+}$ ions with atomic data from the `mom97' \emph{adf04} ADAS source files \cite{ADAS}. As a part of the ionization balance process, the relative numbers of each of these ions making up the total carbon density is calculated. Due to numeric instabilities caused by interpolation limits on the ADAS atomic data, 64 points with temperatures less than 1eV were removed. Each line predicted by the CR model is broadened by convolution with a 1 nm FWHM Gaussian to approximate the DivSPRED spectrometer's instrumental width that is the dominant contributor to the line-widths observed experimentally. The resulting spectra are interpolated onto a 1000-element wavelength array matching the DivSPRED dispersion between 50-160 nm. In total, 1782 separate carbon lines are predicted by the CR model in this wavelength region.\\

Some measure of the dynamic range must be imposed on this dataset to make it more physically reasonable; the lines predicted by the CR model span over 30 orders of magnitude in brightness, a scenario that cannot be replicated in an experimental setting. Spectrometer dynamic ranges are largely set by the bit depth of the detector which is generally in the 12-16 bit range. DivSPRED has a 16 bit (four order of magnitude) detector but also has a four order of magnitude variation in efficiency across the grating resulting in a possible eight order of magnitude difference between the noise floor and the brightness of the brightest line \cite{wood1988}. To reproduce this effect in the simulated dataset, we take the `floor value', $f$ to be $10^{-8}$ and set any point in the dataset that is less than the maximum intensity in the dataset by a factor of $f$ to be $f\times\text{max}(I)$. This replicates the finite observing power of the spectrometer. \\

\begin{figure}[t]
    \centering
    \subfloat{\includegraphics[width=1\columnwidth]{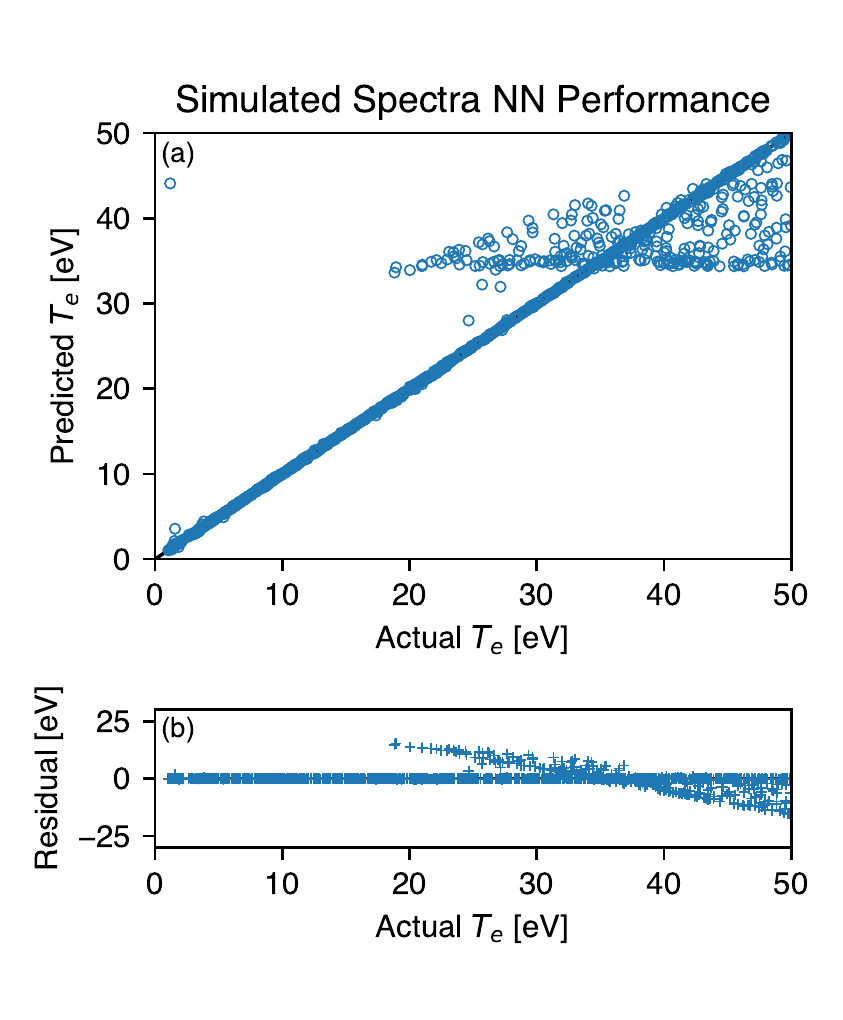}}
    \caption{\label{fig:colradpy_nn_performance} The performance of a NN predicting the electron temperature from simulated spectra produced using a collisional radiative model. The predicted temperatures as a function of the actual temperature in the evaluation test set are shown in (a) alongside the residuals (predicted - measured) in (b). }
\end{figure}

A NN with the same architecture as implemented for the experimental dataset was trained on the simulated spectra dataset (12 layers of 12 neurons, fully connected, feed-forward). It was found that compared with the experimental case, the training of this model benefited from a preprocessing step taking the logarithm of the dataset, a slower rate of learning rate decay ($10\times$ every 200 epochs), a longer criterion for early stopping (50 epochs without improvementI) removal of the $L_2$ regularization and consequently a more gradual training process. The performance of this model is shown in Figure \ref{fig:colradpy_nn_performance} where excellent agreement between the calculated (measured) and predicted temperature is observed for low temperatures. The model tends to under-predict at high temperatures and over-predict at moderate temperatures as was observed in the experimental dataset. The mean absolute residual across the test dataset was 1.5 eV ($\sigma = 3.3$ eV). Below a temperature of 10 eV, the mean absolute error was 0.3 eV ($\sigma=2.8$ eV).\\

\begin{figure}[t]
    \centering
    \subfloat{\includegraphics[width=1\columnwidth]{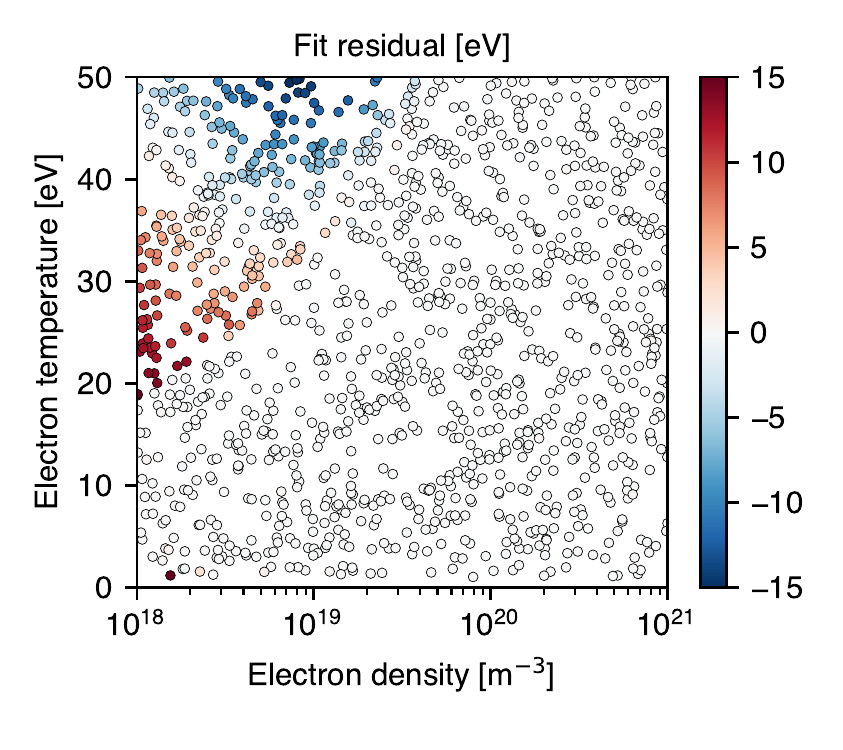}}
    \caption{\label{fig:colradpy_nete_residuals} The difference between real and predicted values as a function of the temperature and density used as inputs for the simulated spectra calculations.}
\end{figure}

\begin{figure*}[t]
    \centering
    \subfloat{\includegraphics[width=1\textwidth]{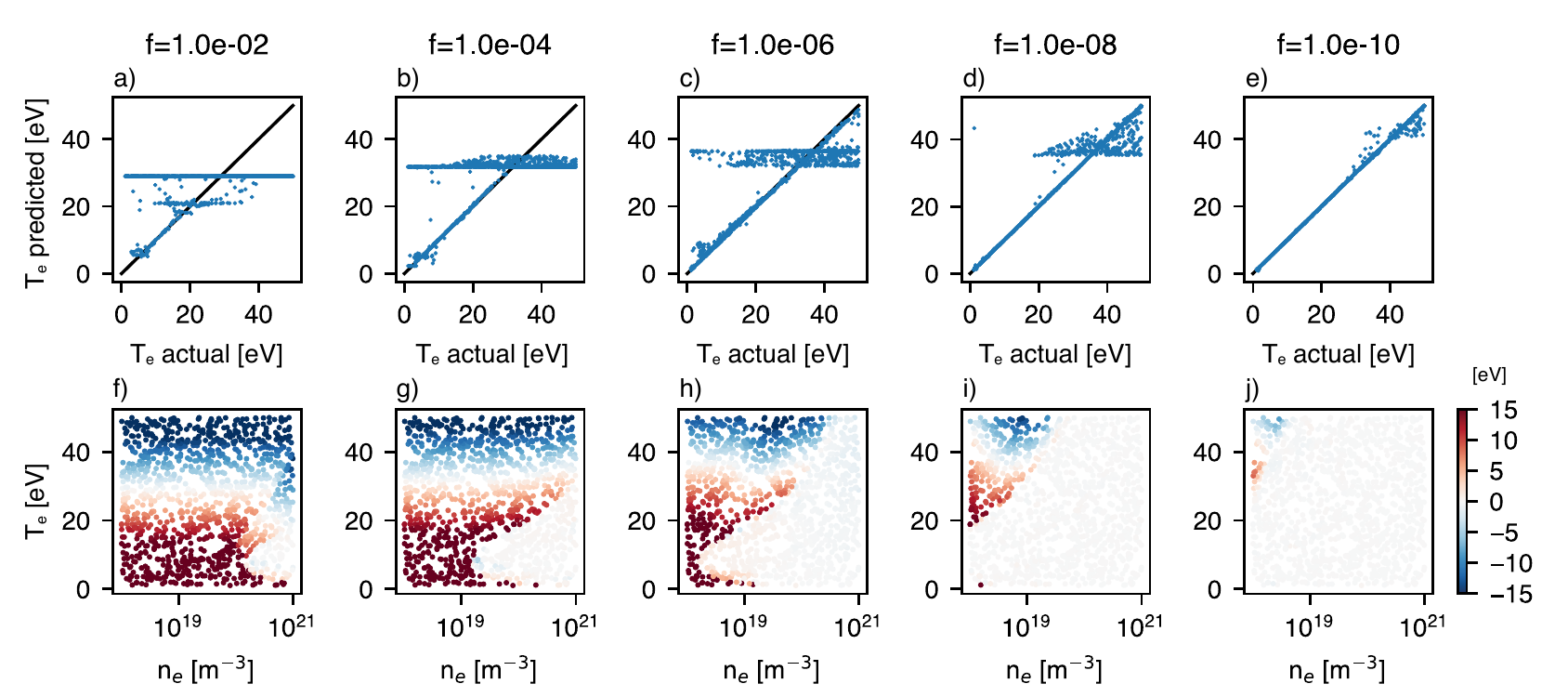}}
    \caption{\label{fig:noise_floor_progression} The NN performance (top row) and distribution of residuals (bottom row) as a function of the noise floor value applied to the simulated spectra dataset, $f$. The highest noise floor, or equivalently the lowest dynamic range, is on the left-side with progressive improvements in each column moving towards the right side.}
\end{figure*}

The relationship between the fit residual and the density and temperature used for the simulated spectra is shown in Figure \ref{fig:colradpy_nete_residuals}. This figure can be compared directly to the equivalent plot for the experimental dataset in Figure \ref{fig:nete_residuals}. The increased coverage of data points in the simulated spectra reveals that it is not just high temperatures that are problematic, but specifically high temperatures at low densities (i.e. only the upper left of the plot). This relationship is explored by examining the dependency of the model performance on the choice of the dynamic range floor value. Figure \ref{fig:noise_floor_progression} gives the NN performance (top row) and the map of evaluation set residuals as a function of density and temperature (bottom row) for a set of values for the noise floor, $f$. Here the noise floor is being reduced (dynamic range increasing) moving from the left- to the right-side columns. Progressively, the NN performance improves as the noise floor reduces with performance improving first at low temperatures and then the high temperature at low densities. Without the limitation of an imposed dynamic range, this NN achieved a mean absolute error of 0.01 eV ($\sigma = 0.01$ eV) across the entire temperature range which amounts to a near-perfect performance.\\

\begin{figure}[t]
    \centering
    \subfloat{\includegraphics[width=1\columnwidth]{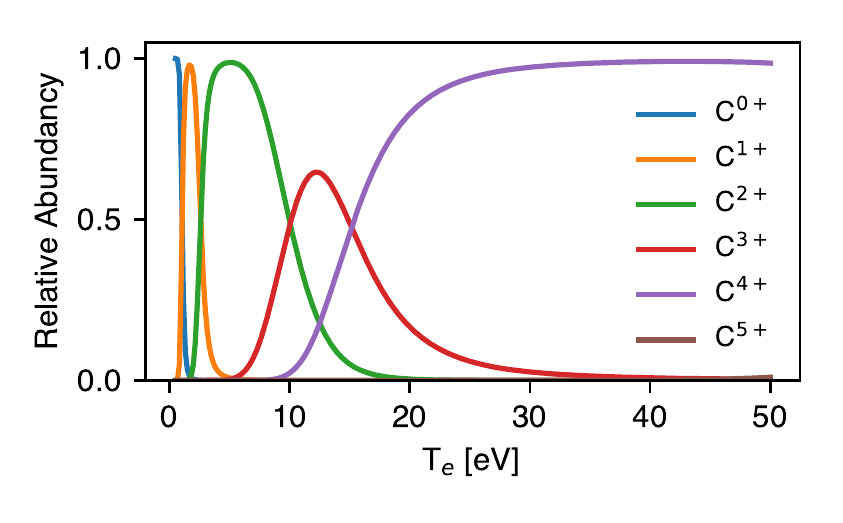}}
    \caption{\label{fig:rel_abund_C} The relative abundance of carbon charge states as a function of electron temperature  at an electron density of $1\times 10^{19}$ \m3. }
\end{figure}

\begin{figure}[t]
    \centering
    \subfloat{\includegraphics[width=1\columnwidth]{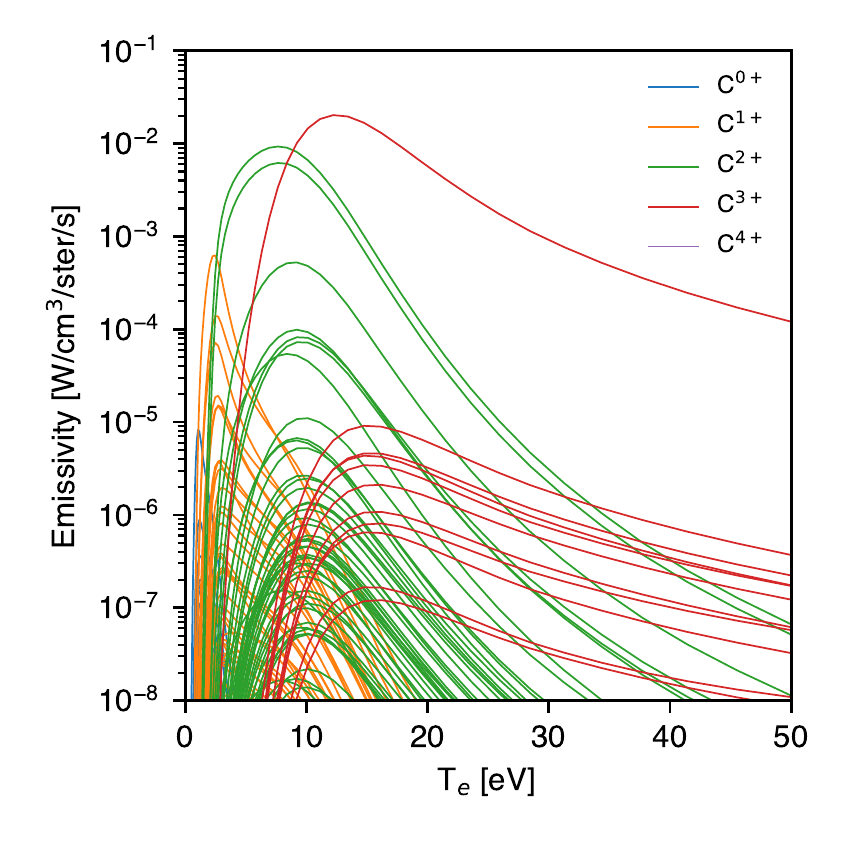}}
    \caption{\label{fig:emissivity_trends}The changing brightness of the simulated carbon lines as a function of electron temperature at $n_e = 1\times10^{19}$ m$^{-3}$. 1782 lines are included encompassing all the predicted emission lines that appear for the wavelengths observable by the DivSPRED spectrometer ($50-160$ nm). }
\end{figure}

The relationship between model performance and the density and temperature can be understood by examining the individual lines and charge-state distribution that determine the simulated spectra. The CR model's ionization balance predicts the relative abundance of each charge state at a given density and temperature. The temperature dependence of this relative abundance is shown in Figure \ref{fig:rel_abund_C}. At the lowest temperature, any carbon present in the plasma is predicted to be 100\% neutral carbon, C$^{0+}$. As the temperature rises, ionization is more likely to occur causing carbon to progressively transition to higher charges states. Above $\sim20$ eV, the carbon population is predominantly C$^{4+}$ which continues until a temperature of around 80 eV when C$^{5+}$ makes up over 50\% of the total carbon. Each individual charge state has a unique set of emission lines that appear and disappear as the charge state abundance rises and falls. This can be seen directly in the emissivity of the lines as a function of temperature as shown in Figure \ref{fig:emissivity_trends}. The charge state abundance translates to a set of lines appearing rapidly and decaying slowly (as a function of temperature) particularly at low temperatures below $\sim 20$ eV. For the NN, this constitutes an information-rich environment whereby a range of features exist whose existence, rather than a particular brightness or ratio of brightnesses, is a strong predictor of particular temperatures. At higher temperatures this is no longer true; there are far fewer lines at higher temperatures and they monotonically decrease with temperature many of which with similar slopes. The collinearity of these lines indicates there is no new information gained for the model to learn from there being multiple lines as they have identical dependencies. Furthermore, there is a large gap in emissivity between the brightest lines and the rest so a limited dynamic range will have a disproportionately negative effect at high temperatures compared to low temperatures where many lines have a similar emissivity magnitude. The emissivity of a single line is not unique with respect to the temperature; the brightness is linearly proportional to the electron density and so there being a limited number of lines at high temperature will be a more substantial effect at a low density where the light is dim and the spectrometer is operating at the bottom end of the dynamic range. The implication of these results is that the difference in NN performance at low and high temperatures observed experimentally can be explained as there being a pronounced difference in the amount of information contained in the spectrum at the low- and high- end of the temperature range which is a natural consequence of the underlying atomic physics that produces the line emission observed by the spectrometer.\\

\begin{figure}[t]
    \centering
    \subfloat{\includegraphics[width=1\columnwidth]{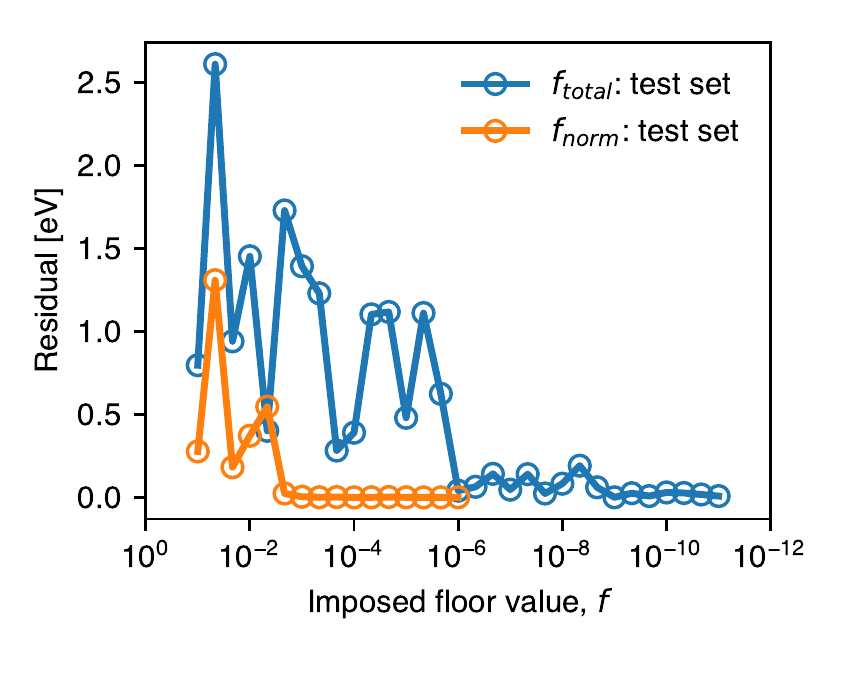}}
    \caption{\label{fig:colradpy_residual_comparison} Model performance as a function of the noise floor value imposed on the dataset before model training. $f_{total}$ corresponds to a floor calculated based on the maximum emissivity of the entire dataset (blue) whereas $f_{norm}$ is calculated for each time slice individually (orange).}
\end{figure}

Requiring a dynamic range in the vicinity of $10^8$ would limit the application of this NN technique to high-end spectrometers and preclude most commercial options that more commonly have dynamic ranges around $10^3-10^4$. However, the implementation of a dynamic range in the simulated spectra assumes that the spectrometer is operated using a static set of settings (e.g. a single gain value) for all density and temperatures which isn't necessarily the case. To test other modes of operation, the simulated dataset was reprocessed by normalizing each individual spectra by the brightest line in that single spectra. In an experimental setting, this equates to continuously tuning the spectrometer's gain and/or integration time to account for the overall brightness of the plasma. While this is often done by an experienced spectrometer operator on a case-by-case basis, it could also be automated in real-time by coupling spectrometer settings to fast photodetector or photomultiplier tube (PMT). \\

A comparison of the two methods for imposing a dynamic range is shown in Figure \ref{fig:colradpy_residual_comparison}. In both cases the residual improves (decreases) as the dynamic range increases, however, the individually normalized spectra ($f_\text{norm}$, orange) decreases substantially faster; a dynamic range of $10^3$ achieves the same excellent performance as the non-normalized dataset at a dynamic range of $10^9$. By normalizing the data in this way, we are removing any absolute intensity information from the spectra; only the relative intensities of the lines are available to the NN as features. It appears that this isn't a detriment to the model training presumably because it is the ratios of lines that are most important for deconvolving temperature and density influence on the line brightness. This approach would have several experimental advantages. Firstly, it may imply that the spectrometers do not need to be routinely absolutely calibrated, a resource-intensive and often difficult process. Secondly, the trained NN is more likely to be generalizable to other machines or divertor geometric configurations since the precise details of light-collection become unimportant. \\

\subsection{Predicting Performance for Nitrogen Spectra in the Visible Spectrum}
 
The collisional radiative model can be used to generate a synthetic spectrum to investigate the scenario of affordable off-the-shelf hardware operated on a metal walled machine. The carbon-based spectra on DIII-D is the result of its carbon walls. However, many tokamaks including ITER and planned next-generation devices are largely envisaged as having metal walls and so will not have any carbon impurities to radiate. In these machines, detachment control is often facilitated by puffing nitrogen gas which radiates effectively in the divertor to induce detachment. A $T_e$ measurement using nitrogen spectra is therefore an attractive option for metal-walled machines. \\

Parameters matching common low-cost low-resolution survey spectrometer in the visible wavelength were used for generating the simulated spectra. The visible wavelength region was chosen to investigate the potential for a low-cost system that would be far easier to implement than a EUV/VUV spectrometer that requires vacuum optics. To this end, a set of 4800 simulated spectra were generated in the same process as outlined for the carbon spectra above. 1350 pixels in the wavelength range of 400 - 850 nm were used to produce a spectrum that was convolved with a Gaussian function with FWHM of 1.4 nm to approximate the spectrometer's instrumental width. Individual spectra were normalized with an imposed dynamic range of $10^3$.  This represents a commercially available spectrometer costing less than \$5000 that can operate at frame rates up to 4.5 kHz. \\

\begin{figure}[t]
    \centering
    \subfloat{\includegraphics[width=1\columnwidth]{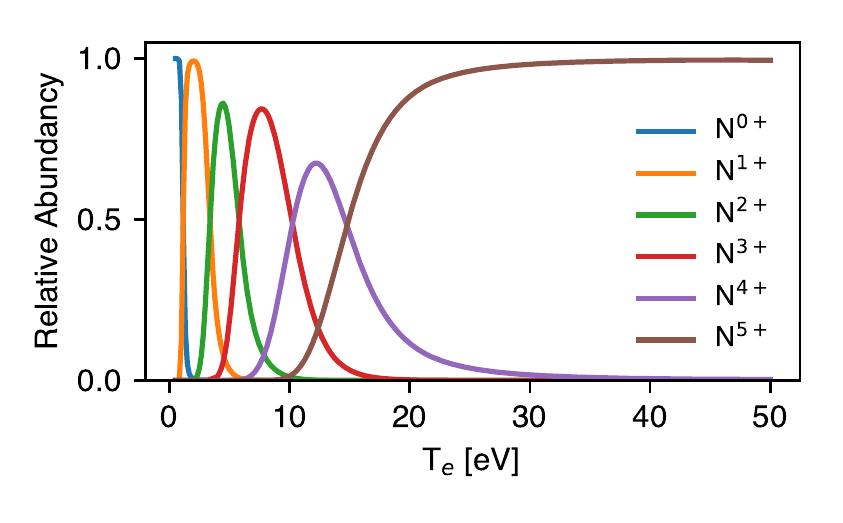}}
    \caption{\label{fig:rel_abund_N} Relative abundance of Nitrogen charge states as a function of temperature at an electron temperature of $n_e = 1\times10^{19}$ \m3.}
\end{figure}

\begin{figure}[t]
    \centering
    \subfloat{\includegraphics[width=1\columnwidth]{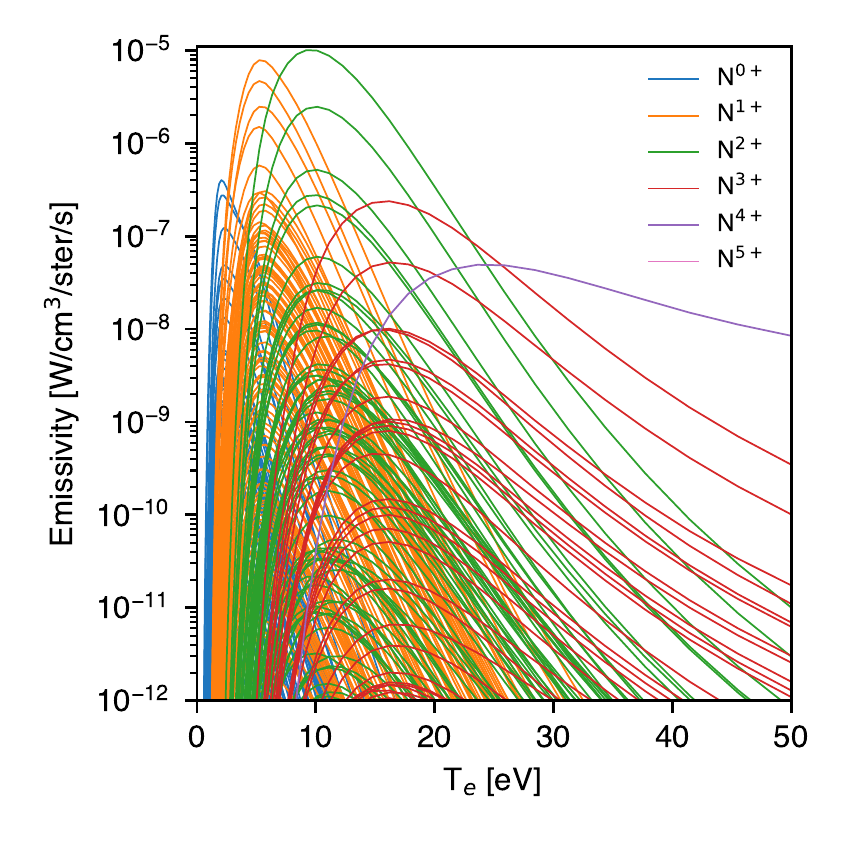}}
    \caption{\label{fig:emissivity_trends_N}The changing brightness of each of the simulated nitrogen lines as a function of electron temperature at $n_e = 1\times 10^{19}$ \m3. 647 lines are included encompassing all the predicted emission lines that appear for the wavelengths observable by a hypothetical affordable spectrometer in the visible wavelengths ($400-850$ nm). }
\end{figure}

The nitrogen emission spectra have many similarities to the carbon spectra owing to the similar atomic number. The charge state abundance of the nitrogen ions as a function of temperature is shown in Figure \ref{fig:rel_abund_N} which can be directly compared to the carbon data in Figure \ref{fig:rel_abund_C}. Nitrogen has one additional charge state compared to carbon and functionally it is placed before the stable helium-like ion and so appears at a temperature below 20 eV. This manifests in the spectra as another set of unique emission lines that will be optimally placed for predicting the temperature of low-temperature plasmas. This can be confirmed by examining the emissivity plot for the 647 nitrogen lines predicted to be in the visible spectrum in Figure \ref{fig:emissivity_trends_N}. While the overall number of lines is lower than were present in the carbon EUV/VUV spectrum, there are a greater number of lines at similarly high emissivities and more transitions between sets of lines (denoted by the colors). This suggests that a ML predictor trained on visible nitrogen lines would perform better than the carbon EUV/VUV data presented in this paper. \\

\begin{figure}[t]
    \centering
    \subfloat{\includegraphics[width=1\columnwidth]{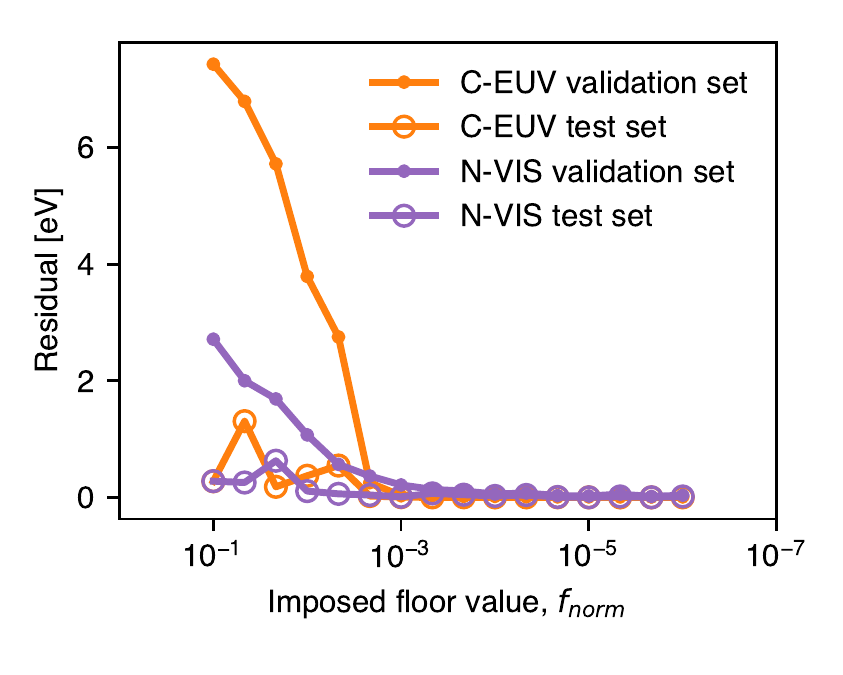}}
    \caption{\label{fig:colradpy_residual_comparison_C_N}Comparison of model residuals as a function of the imposed noise floor value for a NN trained on a simulated set of carbon spectra in the EUV/VUV and a set of nitrogen spectra in the visible wavelengths.}
\end{figure}

The performance of the $T_e$ predictor as a function of imposed floor value (with a lower $f_\text{norm}$ indicating higher dynamic range) is shown in Figure \ref{fig:colradpy_residual_comparison_C_N}. The residual for the validation dataset indicates the relative speed at which the NN trains. The test data indicates the performance of the as-built NN on previously unseen data and confirms the prediction that a NN built on simulated nitrogen visible spectra performs at least as well as one built on carbon VUV/EUV spectra. For a $f_{norm} = 10^{-3}$, the mean absolute error across the 1-50 eV range was 0.1 eV ($\sigma = 0.3$ eV). Below a $T_e$ of 10 eV, the mean error was 0.2 eV ($\sigma = 0.01$ eV) equating to a predicted performance on the same order as DTS itself. To perform the full validation of this proposed system, a synthetic diagnostic could be developed that includes both statistical and neutron-induced noise, molecular and neutral deuterium emission, the potential need to filter bright D$_\alpha$ light, densities high enough to create an optically thick plasma, Zeeman and Stark line splitting, and the relay optics required to position the spectrometer far from the high-radiation environment of fusion devices.\\

\section{Conclusion}
\label{sec:conclusion}

The study in NN performance documented here demonstrates the potential for ML models in augmenting the performance of current fusion diagnostics and presents a pathway for the development of new diagnostic techniques. In the simplest implementation, it shows how a spectrometer combined with a NN can be used to enhance the DTS measurement by increasing the effective acquisition speed at a single point by over an order of magnitude. By operating the two systems simultaneously, the temporal resolution of the measurement is improved while providing a continuous verification of the NN validity (i.e. no model drift). Additionally, the high accuracy and F$_1$ scores of the detachment classifier demonstrates that a combined NN and spectrometer setup is a powerful diagnostic for identifying detachment in its own right. \\

There are multiple avenues by which the performance of the NN trained on experimental data can be improved. The most straight-forward is the generation of a larger and more diverse training dataset; ML performance is closely tied to the data quantity as well as the quality. In particular, the inclusion of both toroidal field directions and data in the presence of impurity puffing is important for producing a model that generalizes well. Adding data from a tangentially-viewing diagnostic such as a filtered camera or fan of spectrometer chords would provide information that the NN could use to learn the role of the position of the brightest plasma volume with respect to the spectrometer's viewing geometry. This may be an avenue by which the single-point predictions described here could be extended to the prediction of a $T_e$ profile. \\

More work on interpreting ML models trained on spectroscopic data could lead to a better understanding and the development of simpler, or perhaps analytical, method for measuring $T_e$ using spectroscopy. It is suspected that information in the continuum emission or the presence of molecular band radiation which only appears at very low temperatures is an important feature that isn't generally accounted for in traditional line-ratio methods for spectroscopic measurement of $T_e$. This information may be the critical component that allows the NN to be trained to predict the $T_e$ at a single point in space using the line-of-sight spectroscopy measurement rather than just an \emph{effective} temperature weighted to the brightest regions of emission. Initial work interpreting the output of a random forests algorithm trained on the peak heights as features indicates that a minimal set of lines can be identified that produce a reliable detachment indicator. In this case, a set of fast filtered detectors such as PMTs could be used to generate a detector that would be well-suited for detachment control. In this way, the training of ML models can contribute to the discovery of new diagnostic techniques. \\

It is envisaged that a NN trained using DTS data may be transferable to other divertor geometries such as the difficult-to-access closed divertor configuration or to other machines that have not implemented a Thomson scattering diagnostic. This could be considered a complex implementation, however, if the NN trained on one machine can be applied to other tokamaks it would enable DTS-like performance for measuring $T_e$ near the strike point and identifying detachment with a simpler hardware setup, increased safety, a smaller footprint and at significantly lower cost compared to DTS. The prediction that the experimental NN demonstrated here in the EUV/VUV is extendable to the nitrogen visible spectra suggests a broad applicability among current and next-generation fusion devices.

\begin{appendix}
  
  \section{Parameterizing Charge Exchange}
  \label{appendix:CX}
\begin{figure}[t]
    \centering
    \subfloat{\includegraphics[width=1\columnwidth]{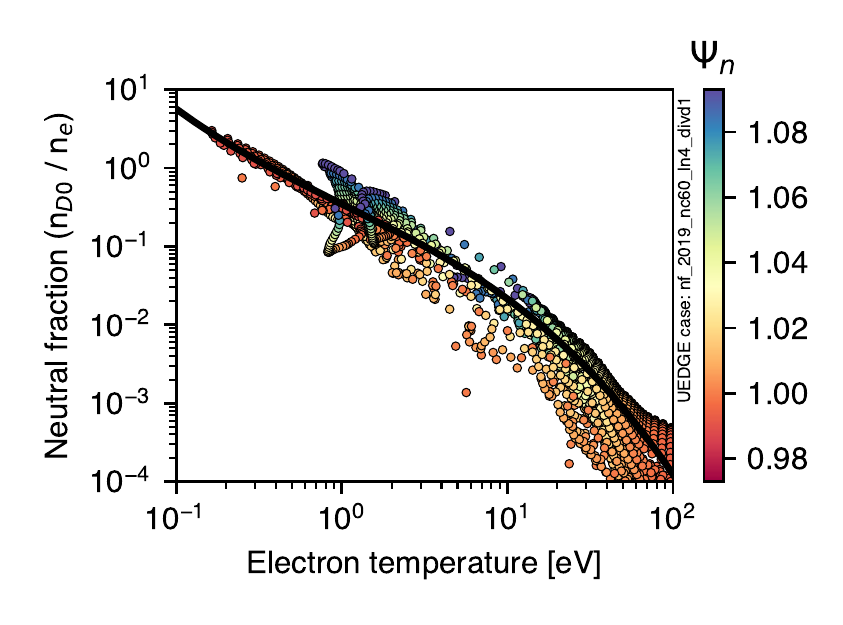}}
    \caption{\label{fig:neutral_fraction} The ratio of neutral atomic deuterium density to electron density in a UEDGE grid cells with $\Psi_n < 1.1$ alongside a heuristic fit.}
\end{figure}

  For inclusion of charge exchange (CX) in collisional radiative modeling, the neutral deuterium density and temperature must be specified alongside $n_e$ and $T_e$ since the CX cross sections depend on them. For this study, the atomic deuterium temperature is taken to be 3.5 eV, the energy of H fragments from electron-impact dissociation of molecules \cite{chabert1998}. While this is a substantial assumption, ColRadPy's emissivity predictions were not overly sensitive to the neutral temperature and the ionization time is very fast in regions where the majority of emission is originating in the divertor. The neutral density is obtained from a fit to the atomic neutral fraction ($n_D / n_e$) predicted by a UEDGE 2D fluid modeling simulation. The UEDGE case used is the `enhanced divertor particle diffusivity' case (\emph{nf\_2019\_nc60\_ln4\_divd1}) from Jaervinen \emph{et al.} \cite{jaervinen2020}. The atomic neutral fraction from each grid cell was calculated excluding $\Psi_n>1.1$ to avoid main-chamber recycling. A trendline was fitted to the functional form:
  \begin{equation}
    \frac{n_D}{n_e} = \exp\left[{a \log(T_e) + b(\log(T_e))^3 + c}\right]
  \end{equation}
  where $a=-1.04$, $b=-0.032$, $c=-1.054$ was found to produce the trendline shown in Figure \ref{fig:neutral_fraction}. This approximation allows the trend of dropping neutral fraction with increasing electron temperature to be included in the CX calculations without the additional complexity of full neutral simulations.\\

\begin{figure}[t]
    \centering
    \subfloat{\includegraphics[width=1\columnwidth]{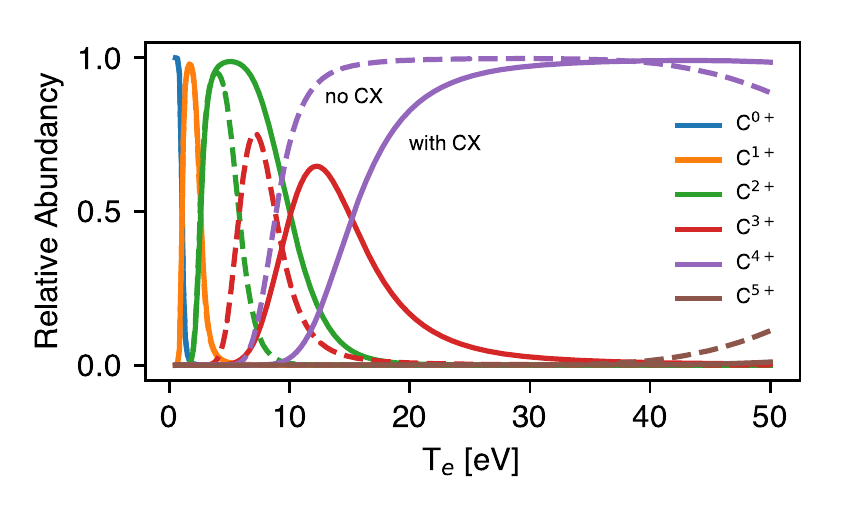}}
    \caption{\label{fig:rel_abund_CX} A comparison of the relative abundancies of carbon ions predicted by ColRadPy at $n_e = 10^{19}$ \m3 with CX turned on (solid lines) and CX turned off (dashed lines).}
\end{figure}

  The net effect of CX inclusion is a movement of the relative abundance of the charge states towards higher $T_e$ and a broadening of their distribution as a function of $T_e$ which better represents the profiles observed experimentally. The effect is shown for carbon ions at an electron density of $n_e = 1\times 10^{19}$ cm$^{-3}$ in Figure \ref{fig:rel_abund_CX} where it can be seen that higher charge states tend to move more than the lower charge states; C and C$^+$ are largely unaffected.

\end{appendix}
\begin{acknowledgments}
The authors would like to thank Jacob West, David Eldon, Colin Chrystal, and Bob Wilcox for excellent advice on many aspects of this work. \\ \newline
The originating developer of ADAS is the JET Joint Undertaking.\\ \newline
This material is based upon work supported by the U.S. Department of Energy, Office of Science, Office of Fusion Energy Sciences, using the DIII-D National Fusion Facility, a DOE Office of Science user facility, under DE-AC52-07NA27344, LLNL LDRD 17-ERD-020, DE-FC02-04ER54698 and DE-SC0015877. LLNL-JRNL-815130.\\ \newline
This report was prepared as an account of work sponsored by an agency of the United States Government. Neither the United States Government nor any agency thereof, nor any of their employees, makes any warranty, express or implied, or assumes any legal liability or responsibility for the accuracy, completeness, or usefulness of any information, apparatus, product, or process disclosed, or represents that its use would not infringe privately owned rights. RLeference herein to any specific commercial product, process, or service by trade name, trademark, manufacturer, or otherwise, does not necessarily constitute or imply its endorsement, recommendation, or favoring by the United States Government or any agency thereof. The views and opinions of authors expressed herein do not necessarily state or reflect those of the United States Government or any agency thereof.

\section*{Data Availability}
Raw data were generated at the DIII-D National Fusion Facility. The data that support the findings of this study are available from the corresponding author upon reasonable request.
\end{acknowledgments}

\end{document}